# Jet acceleration of the fast molecular outflows in the Seyfert galaxy IC5063

C. Tadhunter[1], R. Morganti[2,3], M. Rose[4], J.B.R. Oonk[2], T. Oosterloo[2,3]

**Massive outflows driven by active galactic nuclei (AGN) are widely recognised to play a key role in the evolution of galaxies[1-4], heating the ambient gas, expelling it from the nuclear regions, and thereby affecting the star formation histories of the galaxy bulges. It has been proposed that the powerful jets of relativistic particles launched by some AGN can both accelerate[5-7] and heat[8] the molecular gas, which often dominates the mass budgets of the outflows[5,9]. However, clear evidence for this mechanism in the form of detailed associations between the molecular gas kinematics and features in the radio-emitting jets has been lacking. Here we show that the warm molecular hydrogen gas in the western radio lobe of the Seyfert galaxy IC5063 is moving at high velocities – up to +/-600 km s$^{-1}$ – relative to the galaxy disk. This suggests that the molecules have been accelerated by fast shocks driven into the interstellar medium (ISM) by the expanding radio jets. These results demonstrate the general feasibility of accelerating molecular outflows in fast shocks driven by AGN.**

IC5063 (z=0.0113) is a massive early-type galaxy ($M_* \sim 10^{11}$ $M_\odot$) that hosts both a type II AGN and a powerful double-lobed radio source ($P_{1.4GHz} = 3 \times 10^{23}$ W Hz$^{-1}$). The first signs of AGN-driven outflows in this object were provided by the detection of extended blue wings to the HI 21cm absorption feature and optical [OIII] emission lines at the site of the radio lobe 2.0 arcsec (0.45 kpc) to the west of its nucleus[10,11,12]. Subsequently, a blue wing was also detected in the CO(2-1) emission line profile of the integrated emission from the galaxy, providing evidence for molecular outflows[5]. However, the low spatial resolution of the mm-wavelength CO observations of this and similar objects[5,6,7] prevented a direct link being established between the putative molecular outflows and the relativistic jets and lobes associated with the AGN.

To overcome the resolution problem we have obtained deep, near-infrared long-slit spectroscopic observations of IC5063, taken with the slit aligned along the axis of the extended radio lobes and jets. The observations were made in good seeing conditions (FWHM=0.6 arcsec) and cover the $H_2$1-0S(1)$\lambda$2.128$\mu$m and $H_2$2-1S(2)$\lambda$2.154$\mu$m rotational-vibrational lines of molecular hydrogen, as well as the Br$\gamma$$\lambda$2.166$\mu$m line emitted by the warm ionized gas at the same spatial locations in the galaxy. In Figure 1 we show a grey-scale representation of the long-slit spectrum, as well as line profiles extracted for three key regions in the galaxy. Although extended molecular hydrogen emission is detected along the full 14 arcsec length of the spectroscopic slit, the surface brightness of the emission is particularly high in the regions encompassed by the lobes of the radio source (+/-2 arcsec on either side of the nucleus), consistent with previous HST imaging observations[13]. Most strikingly, the kinematics of the molecular gas are highly disturbed at the position of the western radio lobe, where the $H_2$1-0S(1)$\lambda$2.128μm line shows a broad, complex profile with a full width at zero intensity of FWZI~1,200 km s$^{-1}$; the $H_2$ emission line profile at this location is clearly broader than that of the nucleus or of the eastern radio lobe.

In Figure 2 we show the results obtained by fitting single Gaussian profiles to the $H_2$1-0S(1)$\lambda$2.128μm emission line profile at several spatial locations along the slit. From this it is clear that both the $H_2$ surface brightness and line width peak at the position of the western radio lobe. Moreover, while the molecular gas at large radius follows the rotation curve of the extended disk of the galaxy[10,12], distortions in the radial velocity curve are apparent at the positions of the eastern and western radio lobes. Clearly, the highly disturbed emission line kinematics measured in the radio lobes cannot be explained by the normal gravitational motions of the gas in the galaxy. Therefore these results provide clear and unambiguous evidence that the molecular gas, like the neutral HI gas[10,11], has been accelerated as a result of the interactions between the expanding radio lobes and the ISM in the galaxy disk.

In terms of the comparison with the outflows detected in other phases of the ISM (see Figure 3), the $H_2$ line profile for the western lobe encompasses the full range of blueshifted velocities measured in the broad, HI 21cm absorption line[10,,11], but has a strong, redshifted wing that is not present in the HI feature. The latter difference can be explained by the fact that, whereas the HI absorption line only samples the gas in the foreground of the radio lobe, the $H_2$ emission line samples the outflowing gas moving towards and away from the observer on the near- and far-side of the lobe respectively. In this sense the $H_2$ velocity profile is similar to that of warm ionized gas, as represented by the near-IR Brγ line, whose kinematics closely follow those of the high ionization optical emission lines (e.g. [OIII]$\lambda$5007)[12]. However, although the Brγ velocity profiles cover a similar velocity range to those of the $H_2$ line, they are different in detail (see Figures 1 and 3).

The detection of a weak $H_2$ 2-1S(2)$\lambda$2.154 emission line in the western lobe allows us to estimate the temperature of the molecular gas in the outflow region, since this feature has a higher excitation energy than the $H_2$ 1-0S(1) line. The ratio between the two $H_2$ lines ($H_2$ 2-1S(2)/$H_2$ 1-0S(1) = 0.027+/-0.03) is consistent with a gas temperature of $1913^{+32}_{-68}$ K, assuming that the molecular gas is thermalised. Using this temperature and the spatially integrated $H_2$1-0S(1) luminosity ($L_{H_2} = (1.7 \pm 0.1) \times 10^{32}$ $W$) we estimate[14] a molecular hydrogen mass of $M_{H_2} = (8.2 \pm 1.2) \times 10^2$ solar masses for the western outflow region, which is several orders of magnitude lower than the $H_2$ mass estimated from the blueshifted CO(2-1) emission feature ($2.25 \times 10^7 < M_{H_2} < 1.29 \times 10^8$ solar masses)[5].

Our observations are consistent with a model in which the relativistic jets are expanding through the clumpy ISM in the disk of the galaxy, driving fast shocks into dense molecular clouds embedded in a lower density medium[15,16]. As the molecular gas enters the shocks it is accelerated and simultaneously heated to high temperatures (T > $10^6$ K), ionizing the gas, and dissociating the molecules. The post-shock gas then cools to ~$10^4$ K, emitting emission lines associated with warm ionized gas (e.g. Brγ) as it does so. Further cooling of the gas below $10^4$

K leads to the formation of molecular hydrogen and other molecules, and the near-IR rotational-vibrational lines of $H_2$ are emitted efficiently as the warm gas cools through the temperature range 5,000 – 1,000 K; at this stage there is also sufficient neutral hydrogen gas to allow strong absorption in the HI 21cm line. Eventually the molecular line emission cools the gas to low temperatures (<100 K), where it is detected through the mm-wavelength CO molecular lines. In this scenario, the substantial difference between the $H_2$ masses estimated from the near-IR rotational-vibrational $H_2$ lines and the mm-wavelength CO lines is explained by the fact that the near-IR $H_2$ lines represent a transitory phase in the warm, post-shock gas as it cools, whereas the CO lines represent the total reservoir of gas that has already cooled below 100 K.

Perhaps the greatest uncertainty with this picture concerns how the molecular hydrogen forms in the warm, post-shock gas. Given the high post-shock gas temperatures, it is likely that most, if not all, of the dust present in the precursor gas will be destroyed in the shocks. Therefore the usual mechanism by which $H_2$ forms in the cool ISM of the Milky Way, via catalysis on the surface of dust grains, may not be effective. In this case, it is probable that the $H_2$ formation is catalysed by electrons in the partially ionized cooling gas, via the intermediate formation of $H^-$ ions[17] – the mechanism by which the molecular hydrogen formed in the first proto-stellar cores in the early Universe.

An alternative possibility is that the molecular gas has been accelerated by the slow entrainment and ablation of dense clumps of molecular gas in a hot, post-shock wind. However, in this case it is more difficult to explain the difference between the CO and near-IR $H_2$ mass estimates, because we would expect the high velocity CO and $H_2$ emitting gas to have been heated to the same degree by the entrainment process. Also, we consider it unlikely that the slow entrainment process would occur in the extreme conditions of the western radio lobe, which represents the working surface of the jet.

It is also important to consider why jet-driven molecular outflows like that detected in IC5063 appear to be rare in the general population of nearby Seyfert galaxies[18]. One possibility is that it may be necessary for the jets to collide with high density molecular clouds in the galaxy disks for the phenomenon to be observable: perhaps only the highest density clouds cool sufficiently quickly to avoid being destroyed by their interaction with the lower density, post-shock wind[15]. Although this condition may be met in IC5063, because its jets are propagating in the plane of the disk of the galaxy, in many other Seyfert galaxies the jets and the disks are not co-planar[19].

Overall these results demonstrate the general feasibility of accelerating molecular gas in fast shocks, regardless of whether the shocks are driven by relativistic jets (as in IC5063) or by hot, fast winds originating close to the accretion disks of the AGN[20]. Therefore they are relevant to understanding the acceleration of the massive molecular outflows that have been detected in ultra-luminous infrared galaxies that contain AGN, but lack powerful radio jets[21,22].

Although we cannot entirely rule out the alternative slow entrainment mechanism, we note that if the high velocity molecular gas were indeed formed via cooling in the compressed, post-shock gas, the natural end point of this process would be the formation of stars. Indeed, jet-induced star formation has been invoked to explain the close alignments between the radio and optical/UV structures in high-redshift radio galaxies[23,24,25]. However, it has proved challenging to find definitive evidence for this mechanism, given the presence of

AGN-related continuum components such as scattered quasar light[26] and nebular continuum[27], which are likely to be particularly strong in powerful, high-redshift objects. Currently the best observational evidence for jet-induced star formation is provided by detailed observations of a few well-resolved radio galaxies of relatively low power in the local Universe[28,29]. Clearly, the detection of molecular hydrogen outflows in the western radio lobe of IC5063 lends further credibility to this mechanism.

**METHODS SUMMARY**

The near-IR observations of IC5063 were taken using the medium resolution mode of ISAAC spectrograph on the ESO Very Large Telescope, with the spectroscopic slit aligned along the radio axis (PA295). A standard ABBA nod pattern was employed, with a 20 arcsecond nod throw, 3 arcsecond dither box, and 300s exposures at each position. Four repeats of the basic nod pattern resulted in total exposure time of 4800s, and sky subtraction was affected by subtracting the co-aligned/co-added A and B spectra. The data were then wavelength calibrated using the bright night-sky lines detected in the spectra, and flux calibrated using observations of the B3V star HIP117315 taken at a similar air mass. Use of a 1.0 arcsecond slit resulted in a spectral resolution of R=3000 (100 km s$^{-1}$), and the data cover a useful wavelength range of 2.104 – 2.230μm, with a spatial scale of 0.146 arcseconds per pixel.

Radial velocities, line widths, and line fluxes were determined by using the STARLINK DIPSO package to fit single Gaussian profiles to the emission lines. All the velocities are measured relative to the rest frame of the host galaxy, as determined using the wavelength centroids of the $H_2$ and Brγ emission lines measured in the nucleus (z=0.01131+/-0.00004), and the line widths have been corrected for the instrumental profile.

At the redshift of IC5063, 1.0 arcsec corresponds to 0.224 kpc for our assumed cosmology ($H_0 = 70\ km\,s^{-1} Mpc^{-1}; \Omega_m = 0.3; \Omega_\lambda = 0.7$). Using this cosmology, the stellar mass for IC5063 quoted in the main text was estimated from the 56 arcsecond aperture K-band magnitude[30], assuming a K-band mass to light ratio of $(M/L_K)_\odot = 0.9$.

**References**

1. Fabian, A. Observational evidence of active galactic nuclei feedback. *Ann.Rev.Astron.Astrophys.* **50**, 455 – 489 (2013).
2. Silk, J., & Rees, M. Quasars and galaxy formation. *Astron.Astrophys.* **331**, L1 – L4 (1998).
3. Fabian, A. The obscured growth of massive black holes. *Mon.Not.R.Astron.Soc.* **308**, L39 – L43 (1999).
4. di Matteo, T. et al. Energy input from quasars regulates the growth and activity of black holes and their host galaxies. *Nature* **433**, 604 – 607 (2005).
5. Morganti, R., Frieswijk, W., Oonk, R.J.B., Osterloo, T., Tadhunter, C. Tracing the extreme interplay between radio jets and the ISM in IC 5063. *Astron.Astrophys.* **552**, L4 – L7 (2013).
6. Dasyra, K. M., & Combes, F. Cold and warm molecular gas in the outflow of 4C 12.50. *Astron.Astrophys.* **541**, L7 – L11 (2012).
7. Morganti, R., Fogasy, J., Paragi, Z., Oosterloo, T., Orienti, M. Radio Jets Clearing

the Way Through a Galaxy: Watching Feedback in Action. *Science* **341**, 1082 – 1085 (2013).
8. Guillard, P., et al. Strong molecular hydrogen emission and kinematics of the multiphase gas in radio galaxies with fast jet-driven outflows. *Astrophys.J.* **747**, 95 – 120 (2012).
9. Alatalo, K., et al. Discovery of an active galactic nucleus driven molecular outflow in the early-type galaxy NGC1266. *Astrophys.J.* **735**, 88 – 100 (2011).
10. Morganti, R. et al. A Radio Study of the Seyfert Galaxy IC 5063: Evidence for Fast Gas Outflow. 1998, *Astron.J.* **115**, 915 – 927 (1998).
11. Oosterloo, T. A. et al. A Strong Jet-Cloud Interaction in the Seyfert Galaxy IC 5063: VLBI Observations. *Astron.J.* **119**, 2085 – 2091 (2000).
12. Morganti, R., Holt, J., Saripalli, L., Oosterloo, T.A., Tadhunter, C.N. IC 5063: AGN driven outflow of warm and cold gas. *Astron.Astrophys.* **476**, 735 – 743 (2007).
13. Kulkarni, V. et al. Unveiling the Hidden Nucleus of IC 5063 with NICMOS. *Astrophys.J.* **492**, L121 – L124 (1998).
14. Oonk, J.B.R., Jaffe, W., Bremer, M.N., van Weeren, R.J. The distribution and condition of the warm molecular gas in Abell 2597 and Sersic 159-03. . *Mon.Not.R.Astron.Soc.* **405**, 898 – 932 (2010).
15. Mellema, G., Kurk, J.D., Rottgering, H.J.A. Evolution of clouds in radio galaxy cocoons. *Astron.Astrophys.* **395**, L13 – L16 (2002).
16. Wagner, A.Y., Bicknell, G.V., Umemura, M. Driving Outflows with Relativistic Jets and the Dependence of Active Galactic Nucleus Feedback Efficiency on Interstellar Medium Inhomogeneity. *Astrophys.J.* **757**, 136 – 160 (2012).
17. Hollenbach, D., McKee, C.F. Molecule formation and infrared emission in fast interstellar shocks. I Physical processes. *Astrophys.J.Suppl.Ser.* **41**, 555 – 592 (1979).
18. Riffel, R.A., Storchi-Bergman, T., Winge, C. Feeding versus feedback in AGNs from near-infrared IFU observations: the case of Mrk 79. *Mon.Not.R.Astron.Soc.*, **430**, 2249 – 2261 (2013)
19. Schmitt, H.R., Donley, J.L., Antonucci, R.R.J., Hutchings, J.B., Kinney, A.L. A Hubble Space Telescope Survey of Extended [OIII] Emission in a Far-Infrared selected sample of Seyfert Galaxies: Observations. *Astrophys.J.Suppl.Ser* **148**, 327 (2003).
20. Zubovas, K., King, A. Galaxy-wide outflows: cold gas and star formation at high speeds. *Mon.Not.R.Astron.Soc.* in press, arXiv:1401.0392 (2014).
21. Veilleux, S. et al. Fast molecular outflows in luminous galaxy mergers: evidence for quasar feedback from Herschel. *Astrophys.J.* **776**, 27 – 48 (2013).
22. Cicone, C. et al. Massive molecular outflows and evidence for feedback from CO observations. *Astron.Astrophys..* **562**, 21 – 46 (2014).
23. McCarthy, P. J., van Breugel, W., Spinrad, H. & Djorgovski, S. A correlation between the radio and optical morphologies of distant 3CR radio galaxies. *Astrophys.J.* **321**, L29 – L33 (1987).
24. Rees, M. J. The radio/optical alignment of high-z radio galaxies - Triggering of star formation in radio lobes. *Mon.Not.R.Astron.Soc.* **239**, P1 – P4 (1989).
25. Gaibler, V., Khochfar, S., Krause, M., Silk, J. Jet-induced star formation in gas-rich galaxies. *Mon.Not.R.Astron.Soc.* **425**, 438 (2012).
26. di Serego Alighieri, S., Fosbury, R.A.E., Tadhunter, C.N., Quinn, P.J. Polarized light in high-redshift radio galaxies. *Nature* **341**, 307 – 309 (1989).

27. Dickson, R., Tadhunter, C., Shaw, M., Clark, N., Morganti, R. The nebular contribution to the extended UV continua of powerful radio galaxies. *Mon.Not.R.Astron.Soc.* **273**, L29 – L33 (1995).
28. Croft, S. et al. Minkowski's Object: A Starburst Triggered by a Radio Jet, Revisited. *Astron.J.* **647**, 1040 – 1055 (2006).
29. Crockett, R., et al. Triggered star formation in the inner filament of Centaurus A. *Mon.Not.R.Astron.Soc.,* **421**, 1603 – 1623 (2012).
30. Griersmith. D., Hyland, A.R., Jones, T.J. Photometric properties of bright early-type spiral galaxies. IV – Multiaperture UBVJHK photometry for the inner/bulge regions of 65 galaxies. *Astron.J.*, 1982, 1106 – 1126 (1982).

**Acknowledgements.** Based on observations collected at the European Southern Observatory, Chile (program: 290.B-5162). C.T. and M.R. acknowledge financial support from the UK Science and Technology Research Council. R.M. gratefully acknowledges support from the European Research Council under the European Union's Seventh Framework Programme (FP/2007-2013) /ERC Advanced Grant RADIOLIFE-320745.

**Author contributions.** C.T. and R.M. led the project and the scientific interpretation of the data, and C.T. wrote the text of the paper. M.R. reduced the near-IR spectroscopic data. R.O. and T.O. contributed equally to the analysis and interpretation of the results.

**Author Information.** Correspondence and requests for materials should be addressed to C.T. (c.tadhunter@sheffield.ac.uk).

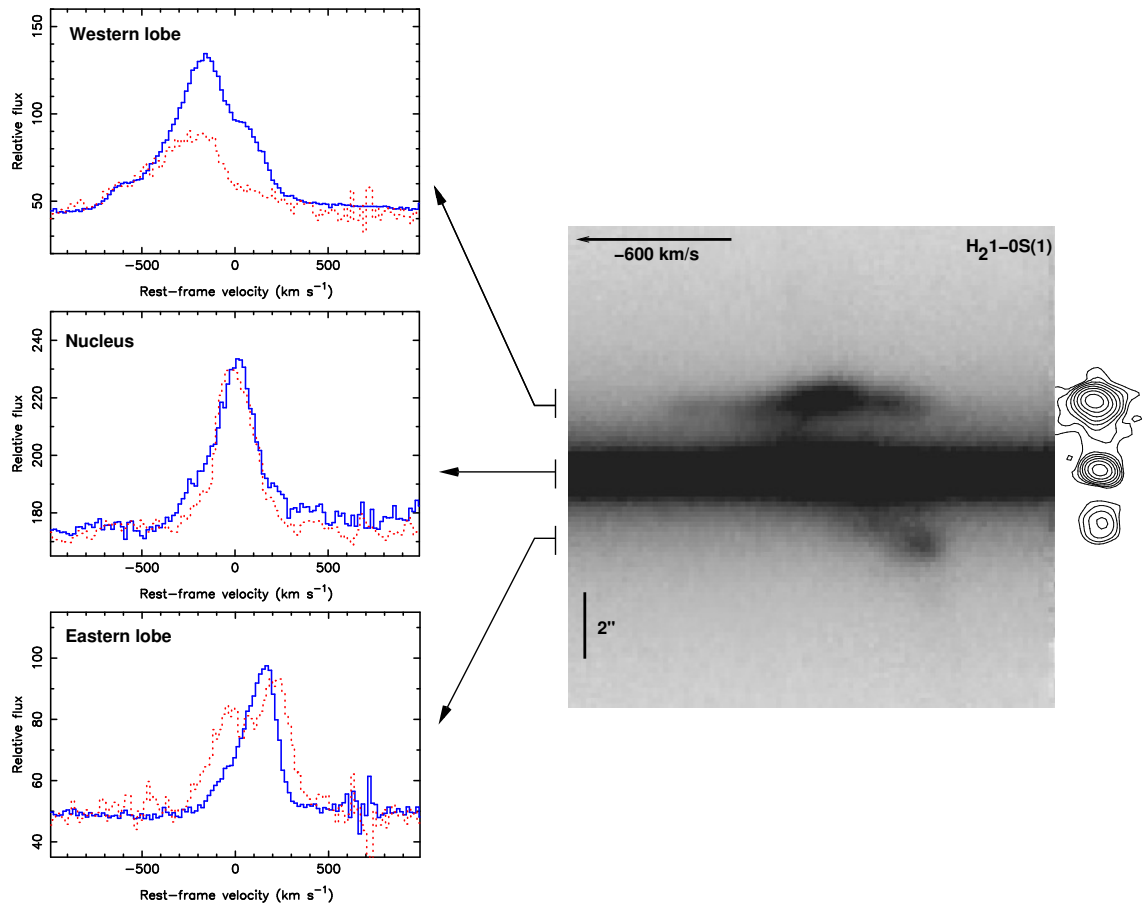

**Figure 1. Signs of extreme kinematic disturbance in the western radio lobe of IC5063.** The central panel shows a grey-scale representation of our long-slit, near-IR (K-band) spectrum of IC5063, covering a wavelength range centred on the H$_2$1-0S(1) line. For comparison, a scaled version of the 1.4GHz radio map of the source is presented to the right. The velocity profiles derived from spectra extracted from three spatial locations across the galaxy are presented to the left, where the solid blue lines represent the H$_2$1-0S(1)$\lambda$2.128 feature, and the dotted red lines represent the Br$\gamma$ feature.

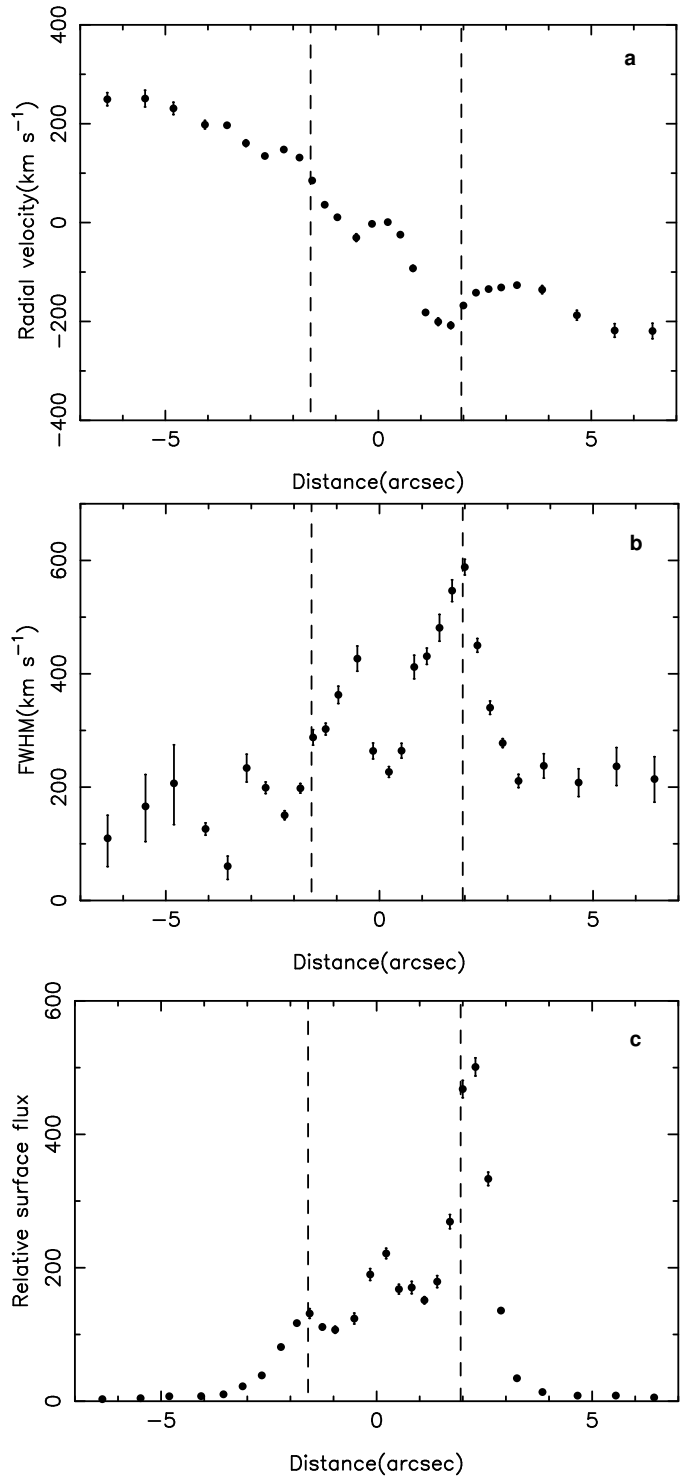

Figure 2. Spatial variations in $H_2$ 1-0S(1) emission line properties. The different panels show the variations in (a) radial velocity, (b) line width (FWHM), and (c) relative surface flux of $H_2$1-0S(1)$\lambda$2.128mm as a function of position along the slit. Distances are measured relative to the centroid of the galaxy continuum emission, and the dashed vertical lines indicate the positions of the centroids of the radio lobes (east to the left, and west to the right). The error bars reflect the one sigma uncertainties in emission line properties derived from single Gaussian fits to the line profiles.

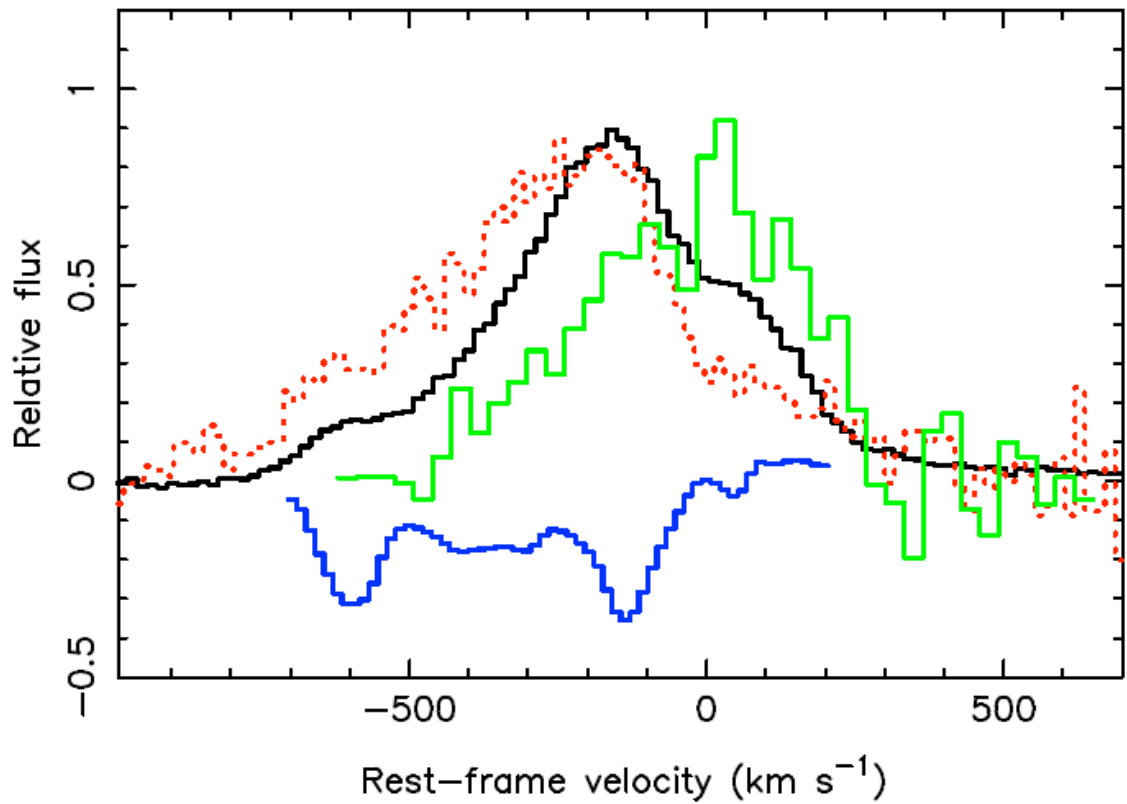

**Figure 3. The multiphase outflow in IC5063.** $H_2$1-0S(1) (black, solid) and Brγ (red, dotted) velocity profiles for the western lobe of IC5063 are compared with the spatially-integrated HI 21cm absorption[10] (blue, solid) and CO(2-1)[5] (green, solid) velocity profiles. The flux scaling between the different profiles is arbitrary.